\documentclass[12pt,a4paper]{article}

\usepackage{times} % Times New Roman Font
\usepackage[a4paper, lmargin=2cm, rmargin=2cm , tmargin=2.5cm , bmargin=2.25cm  ,footskip=1.25cm]{geometry} % To set the margins
\usepackage{graphicx} 
\usepackage[square,sort,comma,numbers]{natbib} % Required to change bibliography style
\usepackage{amsmath} % Required for some math elements
\usepackage[hyphens]{url} % To cite URLs
\usepackage{soul}
\usepackage{fancyhdr} % To use the image as header
\usepackage[explicit]{titlesec} % To format the titles
\usepackage{abstract} % To format the abstract
\usepackage{authblk} % To format the authors list with numbers and affiliations
\usepackage{tabularx}
\usepackage{multirow} % To have multi rows
\usepackage{colortbl} % To color table lines
\usepackage{booktabs}
\usepackage{array}
\usepackage[super]{nth} % for cardinal numbers typesetting
\usepackage{algorithm}
\usepackage[end]{algpseudocode}
\usepackage{hyperref}
\usepackage{numprint}
\usepackage{subcaption}

\newcommand{\etal}{\textit{et al.}}
\titleformat{\section}{\normalfont\bfseries}{\thesection}{1em}{\MakeUppercase{#1}}  % section bold face upper case

\titleformat{\subsection}{\normalfont\bfseries\small}{\thesubsection}{1em}{{#1}} % subsection bold face lower case

\titleformat{\subsubsection}{\normalfont\small}{\thesubsubsection}{1em}{{#1}} % subsubsection normal lower case

\algrenewcommand\algorithmicend{\textbf{end}}
\algrenewtext{EndFor}{\algorithmicend}

\makeatletter
\newcommand{\thickhline}{ \noalign {\ifnum 0=`} \fi \hrule height 1pt \futurelet \reserved@a \@xhline }
\newcommand{\morethickhline}{ \noalign {\ifnum 0=`} \fi \hrule height 2pt \futurelet \reserved@a \@xhline }
\newcolumntype{"}{@{\hskip\tabcolsep\vrule width 1pt\hskip\tabcolsep}}
\makeatother

\newlength{\Oldarrayrulewidth}

\def\etal{\emph{et al.}}

\begin{document}

%-------- TITLE AND AUTHORS --------%
\newcommand{\gomez}{G\'{o}mez} %for when your name has these great features
\title{\normalsize\normalfont\bfseries \MakeUppercase { Study of the asteroid Bennu using geodesyANNs and Osiris-Rex data } }
\date{}
\author[(1)]{Moritz von Looz}
\author[(2)]{Pablo \gomez}
\author[(3)]{Dario Izzo}
%\author[4]{}
%\author[5]{}
\affil[(1)]{ \textit{European Space Agency, Keplerlaan 1, 2201 AZ Noordwijk, moritz.von.looz@esa.int}}
\affil[(2)]{ \textit{European Space Agency, Keplerlaan 1, 2201 AZ Noordwijk, dario.izzo@esa.int}}
\affil[(3)]{ \textit{European Space Agency, Keplerlaan 1, 2201 AZ Noordwijk, pablo.gomez@esa.int}}
%\affil[4]{}
%\affil[5]{}

\renewcommand\Authands{, }
\renewcommand{\Authfont}{\normalsize\normalfont \bfseries}
\renewcommand{\Affilfont}{\normalsize\normalfont}
\renewcommand{\abstractnamefont}{\bfseries\normalsize\MakeUppercase} % makes the abstract title uppercase and bold faced

\maketitle
%-------- END OF TITLE AND AUTHORS --------%

\begin{abstract}
Asteroids and other small bodies in the solar system tend to have irregular shapes, owing to their low gravity. This irregularity does not only apply to the topology, but also to the underlying geology, potentially containing regions of different densities and materials. The topology can be derived from optical observations, while the mass density distribution of an object is only observable, to some extent, in its gravitational field.

In a companion paper, we presented geodesyNets, a neural network approach to infer the mass density distribution of an object from measurements of its gravitational field. In the present work, we apply this approach to the asteroid Bennu using real data from the Osiris Rex mission. The mission measured the trajectories of not only the Osiris Rex spacecraft itself, but also of numerous pebble-sized rock particles which temporarily orbited Bennu.

From these trajectory data, we obtain a representation of Bennu's mass density and validate it by propagating, in the resulting gravity field, multiple pebbles not used in the training process. The performance is comparable to that of a polyhedral gravity model of uniform density, but does not require a shape model. As little additional information is needed, we see this as a step towards autonomous on-board inversion of gravitational fields.
\end{abstract}

\thispagestyle{fancy}

\section{INTRODUCTION}
Recent missions to small asteroids, such as the Osiris-rex mission~\cite{lauretta2017osiris} to the asteroid (101955) Bennu, the Rosetta mission~\cite{glassmeier2007rosetta} to comet Churyumov–Gerasimenko or the Hayabusa mission~\cite{fujiwara2006rubble} to (25143) Itokawa provide a wealth of new insights into the composition and structure of their target bodies. Among these is the heterogeneity of mass distributions which can be significant even of small bodies which did not have sufficient mass to undergo internal differentiation. Itokawa is thought to be such a case, consisting of two parts having different densities ~\cite{fujiwara2006rubble}, while Bennu has a lower density in its equatorial bulge~\cite{Scheereseabc3350}.
Each new observation yield scientific insights into their history -- Itokawa is considered to be the unique result of a collision of two bodies of different densities. In addition to its scientific benefit, modeling an object's gravitational field is crucial for orbit planning, an additional challenge around heterogeneous bodies.

Apart from rotational observations~\cite{waszczak2015asteroid}, the main approach to gain insights about an object internal mass distribution is by observation of its gravitational field. Two common methods to model deviations from the gravity field produced by a uniformly homogeneous sphere are the use of spherical harmonics~\cite{eshagh2020satellite} and polyhedral models~\cite{paul1974gravity}. %TODO: not quite happy with that sentence yet

In contrast to these approaches, we employ a neural network architecture we call \emph{geodesyNets} (see Section~\ref{sec:methods}) to directly learn the bodies' mass density function. In our companion paper~\cite{izzo2021geodesy} we describe the architecture of geodesyNets and experiments on synthetic data, which show an error of under 1\% even when close to the asteroids' surface. This, however, assumes observations of the experienced acceleration on a grid of measurement points achieving full surface coverage and avoid of any unmodelled noise.

In the present work, we validate our approach on real trajectory data derived from the Osiris Rex mission, deriving the accelerations by numerical differencing of observed velocities of orbiting pebbles.

Challenges in this approach include measurement noise inherent to any observational study, as well as additional non-gravity forces, for example pressures from solar or thermal radiation. In addition to the Osiris-Rex spacecraft, the asteroid Bennu is occasionally orbited by small, pebble-sized particles. These are likely emitted in thermal fracturing events~\cite{Laurettaeaay3544} and often orbit the asteroid a few times before either re-impacting or escaping, thus giving a rich, albeit indirect, and detailed view of the local gravity field. However, due to their small size and thus large area-to-mass ratio, non-gravity forces have an even larger effect on these particles. The pebble trajectories are provided in a dataset published by Scheeres \etal~\cite{Scheereseabc3350}. %TODO: check that Scheeres et al. actually were the ones that created them.

We consider two scenarios: In a first case we name \emph{non-differential}, we train our network to predict the mass density function of the body without requiring a shape model.\footnote{Note that Bennu, like many asteroids, rotates. Thus it is still necessary to measure the rotation and establish a body-fixed frame of reference.} In a second case we refer to as \emph{differential}, we assume that a shape model is available and we may then model the differences in the mass density distribution with respect to a homogeneous density, hoping to explore in more details the object's heterogeneity. To validate the fidelity of the resulting gravitational field, we propagate the orbits of several particles based on the derived gravitational field. In this \emph{orbit propagation} step, we compare the propagated trajectories based on the neural density field with those actually flown by the pebbles as reported by Scheeres \etal~\cite{Scheereseabc3350}. Note that since their reported pebble trajectories are orbital fits derived from optical measurements, they \emph{already} include modeling of the balance of forces, among them a spherical harmonics modeling of the gravitational heterogeneity, solar radiation pressure and a detailed thermal model of Bennu itself. As we omit a full thermal model of Bennu, some of these modeled forces will appear as noise in our experiments.

Our experiments (Section~\ref{sec:experiments}) show that our model infers the shape of Bennu from the measured accelerations even in the presence of unmodelled forces, which is demonstrated in the ability to propagate trajectories even in the absence of a shape model.

To ease reproducibility, we provide access to the code used to perform these experiments at \href{https://github.com/mlooz/bennu-sampler/}.

\section{RELATED WORK}

\subsection{Gravitational Modeling of Small Bodies}
%spherical harmonics

%polyhedral models

Scheeres \etal~\cite{Scheereseabc3350} combine polyhedral models and spherical harmonics in what they call a \emph{split gravity} model to mitigate convergence issues of spherical harmonics below the circumscribing sphere of the body in question.

McMahon \etal~\cite{mcmahon2020dynamical} simulate trajectories around Bennu with this force model and derive that many simulated trajectories re-impact at Bennu's equator, leading to equatorial fill-in.

\section{METHODS}
\label{sec:methods}
We train a neural network to estimate the mass density $f: (x,y,z) \rightarrow \rho$ within the body in question. As part of this process, we evaluate the network at a large number of points, then integrate over these to calculate the resulting gravitational field. On target points \emph{outside} of the body, we compare the gravitational acceleration predicted via the network with the real acceleration from either synthetic or real-world data. The difference is normalized to form a loss function suitable for training a geodesyNet. We perform this process until the mass distribution $f$ converges to represent the resulting gravitational field of the body so that it matches the measurements as well as possible. This approach is presented in detail, together with results on synthetic data, in our companion paper~\cite{izzo2021geodesy}.

\subsection{Network}
For the network itself, we use densely connected layers in combination with a sinus activation function, similar to the SIREN networks of Sitzmann et. al.~\cite{sitzmann2020implicit}. This leads to a complete differentiability of the network, which facilitates training.

We distinguish between the \emph{non-differential} and \emph{differential} cases.  In cases where a shape model is available, a higher accuracy can be reached by learning the \emph{differences} of mass density compared to a uniform density model.

Given a batch of size $n$ the loss function in the non-differential training is a modified mean absolute error (MAE) computed on the accelerations $\hat y_i, i=1..n$ induced by the mass density function the network predicted compared to ground-truth observations $y_i, i=1..n$. Given the normalization factor $\kappa$ - as in the companion paper~\cite{izzo2021geodesy} - the loss is then

\begin{equation}
\mathcal L_{\kappa MAE} = \frac  1n \sum_{i=1}^n{\left|y_i - \kappa \hat y_i\right|} = \frac  1n \sum_{i=1}^n{\left|y_i - \frac{\sum_{i=1}^n \hat y_i y_i}{\sum_{i=1}^n y_i^2} \hat y_i\right|}.
\label{eq:network-loss}
\end{equation}

In the differential training the network is tasked to infer the density difference compared to a homogeneous density distribution represented by a mascon model. The ground-truth accelerations $y_i$ are once again the observations. The label, however, is the sum of the acceleration $y^{m}_i$ induced by a homogeneous body modeled by a mascon model and the one caused by the network prediction $\hat{y_i}$. Thus, the total loss in the differential training is %TODO

\begin{equation}
    \mathcal L_{\kappa MAE} = \frac  1n \sum_{i=1}^n \left| y_i - y^{m}_i - \kappa \hat{y_i} \right|.
\end{equation}

In all the following experiments, we use a fully connected network with 9 layers of 100 neurons each, three input neurons (for the coordinates) and one output neuron. For non-differential training, we use sigmoid or absolute activation functions and tanh for differential training.

We train for 10000 iterations with a batch size of 100 samples. The training uses an ADAM optimizer with an initial learning rate of $1^{-4}$.  A learning rate schedule is used to decay the learning rate up to a minimum learning rate of $1^{-6}$. The necessary numerical integrations were computed using a vectorized version of the composite trapezoidal rule. The acceleration was computed as \begin{equation}
\label{eq:quadrature}
   \mathbf a(\mathbf r) = G \int_{\mathbf x \in V} \frac{\rho(\mathbf x)}{|\mathbf r - \mathbf x|^3}\left(\mathbf r - \mathbf x\right) dV,
\end{equation} 
where $G$ is the Cavendish constant, $x$ the points used to sample the density $\rho(x)$ from the network, $r$ the observation point and $V$ is a cubical domain around the body.

\section{DATA}
\label{sec:data}
To validate our approach on real data, we use trajectory data from the Osiris-Rex mission, provided in the form of a SPICE~\cite{ACTON20189} kernel by the Osiris Rex science team. The dataset published by Scheeres et. al.~\cite{Scheereseabc3350} additionally provides estimated trajectories for small rocks ejected from Bennu itself. The trajectory of the spacecraft can be measured with higher precision, while the rock ejecta (also called \emph{pebbles}) are more numerous and come closer to Bennu's surface, where the non-homogeneity of Bennu's gravity field is expressed more strongly.

From these trajectories, we obtain samples of the acceleration by numerical differentiation of the velocities with a five-point stencil and a step size of 0.1 seconds. As Bennu is rotating, we compute the accelerations in an inertial frame and then rotate them into a body-fixed frame before passing them to the neural network.

\subsection{Osiris Rex}
The Osiris Rex spacecraft entered Bennu's orbit on December 3rd, 2018 and stayed there until May 2021. The periapsis of its orbit was decreased down to 700m, before it approached Bennu's surface for the Touch-And-Go \emph{(TAG)} mission. The main source of accelerations that are not due to Bennu's gravity are active maneuvers of the craft, fortunately these are orders of magnitude larger and thus easy to recognize. Figure~\ref{fig:osiris-rex-points} shows an illustration of samples from the trajectory of Osiris-Rex. 

\subsection{Rock Ejecta}
\begin{figure}
\centering
\subcaptionbox{Points from trajectory of Osiris-Rex\label{fig:osiris-rex-points}}
{\includegraphics[width=0.45\linewidth]{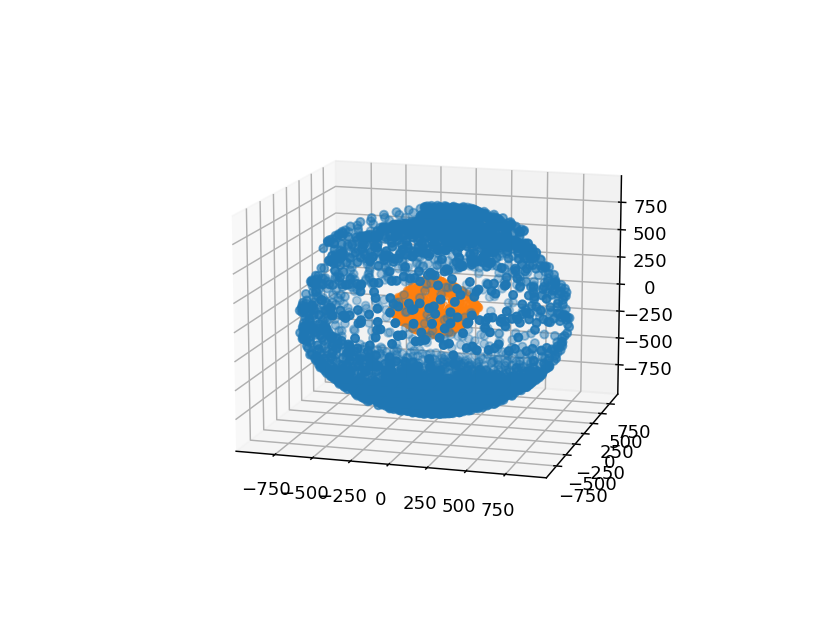}}
\subcaptionbox{Trajectory of a single pebble\label{fig:one-pebble}}
{\includegraphics[width=0.45\linewidth]{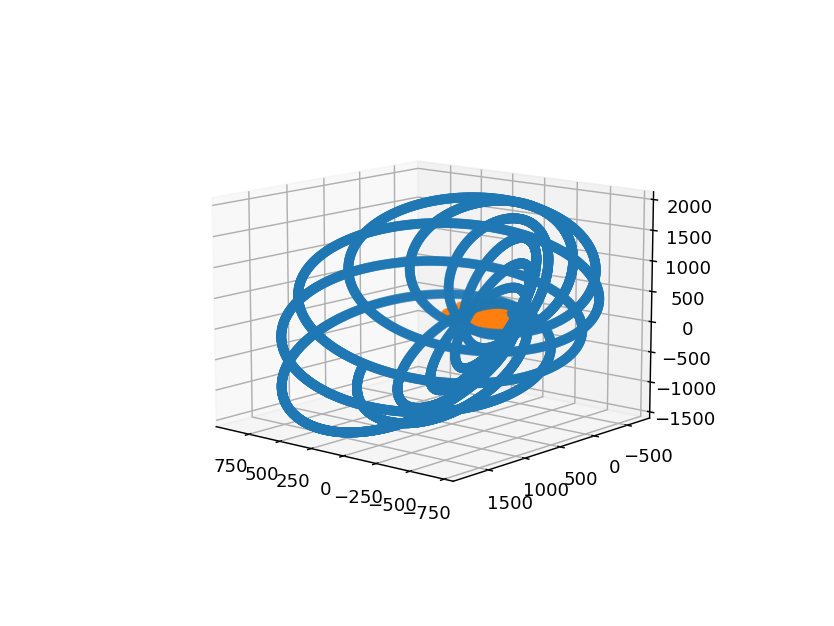}}
\caption{Illustration of Bennu (orange) and points sampled from the trajectories of Osiris-Rex and a single pebble, in inertial frame}
\label{fig:sample-points}
\end{figure}

During the time of observation, Bennu ejected multiple small, pebble-sized rock particles. Many of these stayed in orbit around Bennu for several days before either falling back to Bennu's surface or escaping its gravitational influence. Their trajectories yield additional samples of the gravity field at close distance. However, due to their small size, they have a high surface-to-mass ratio and radiation effects play a larger role. An illustration of such a trajectory is shown in Figure~\ref{fig:one-pebble}. Due to their large number and close proximity, we focus exclusively on samples from pebble trajectories in our experiments.

We consider the 313 pebbles catalogued by Scheeres \etal~\cite{Scheereseabc3350} and sample positions and accelerations from them in regular intervals of a few minutes. The time each pebble stays in orbit around Bennu before re-impacting or escaping ranges from a few hours up to 14 days.

\subsection{Solar Radiation Pressure and Other Forces}
\label{sec:srp}
For those positions that are not shadowed by Bennu itself, we estimate the solar radiation pressure. At Bennu's distance from the sun, the solar radiation emits a pressure of about 4.6 micropascal, and assuming spherical particles with an average diameter of 1 cm and a density of 2 $g/m^2$ they experience a resulting acceleration of around $4\times 10^{-10} m/s^2$, which matches with those reported by Scheeres \etal~\cite{Scheereseabc3350}. We subtract the acceleration caused by solar radiation pressure from the data used for training the neural network.

%Even afterwards, there is a large amount of non-gravitational forces, we thus calculate for each sample the acceleration that would be expected if Bennu were a uniform sphere, and remove all samples whose acceleration deviates more than 15\% from this.

\begin{figure}
\includegraphics[width=0.23\linewidth]{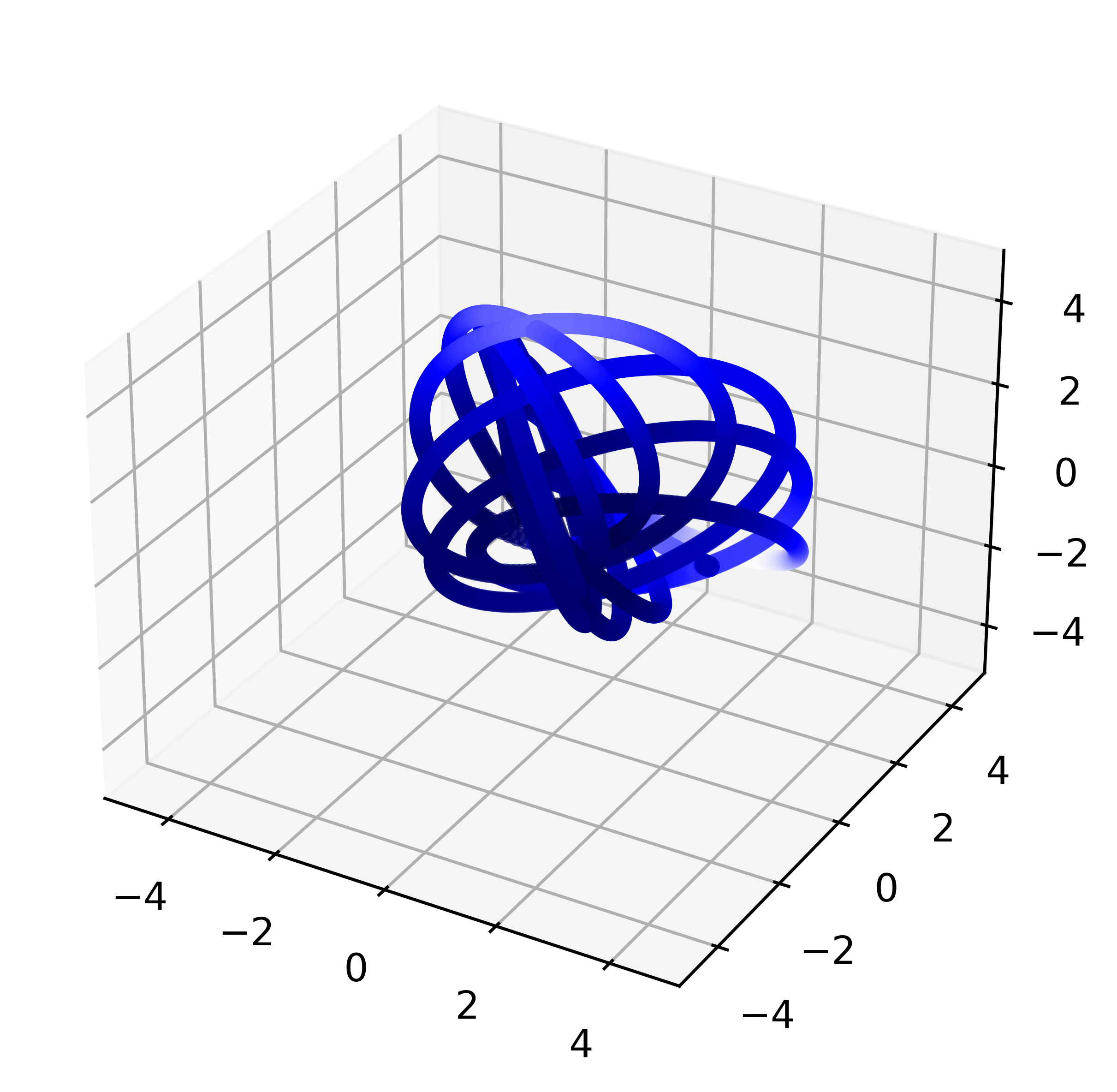}\quad
\includegraphics[width=0.23\linewidth]{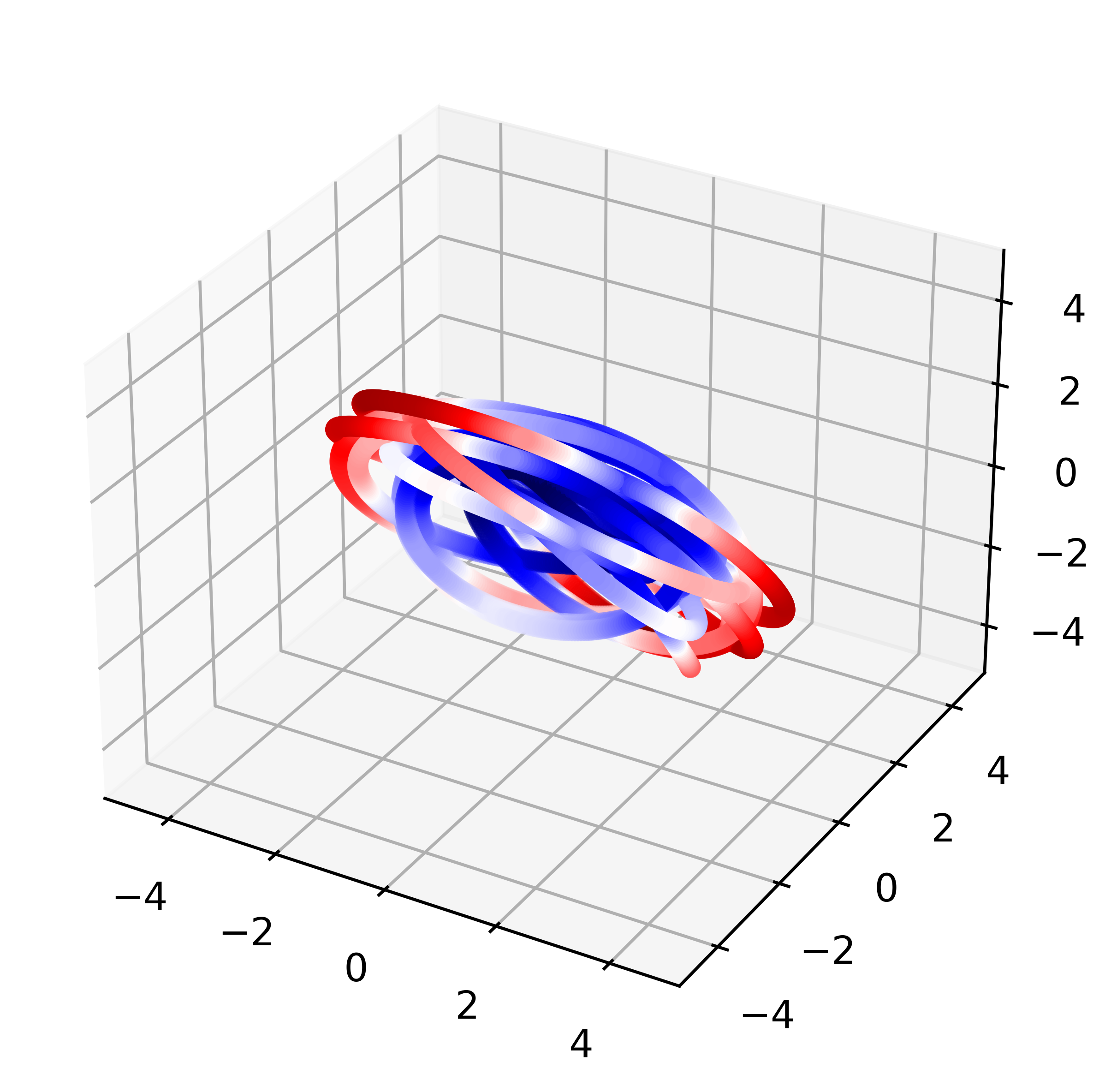}\quad
\includegraphics[width=0.23\linewidth]{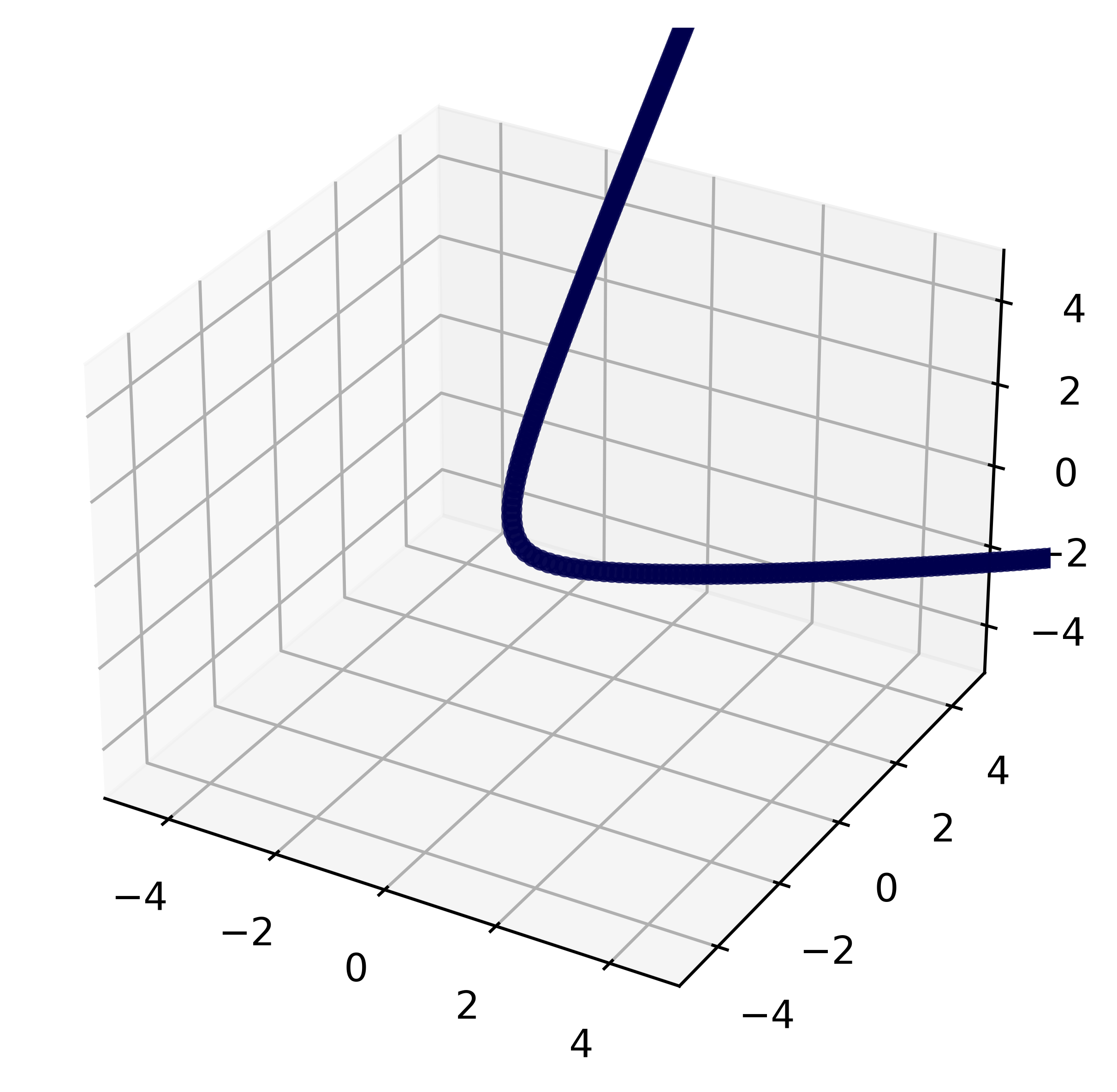}\quad
\includegraphics[width=0.23\linewidth]{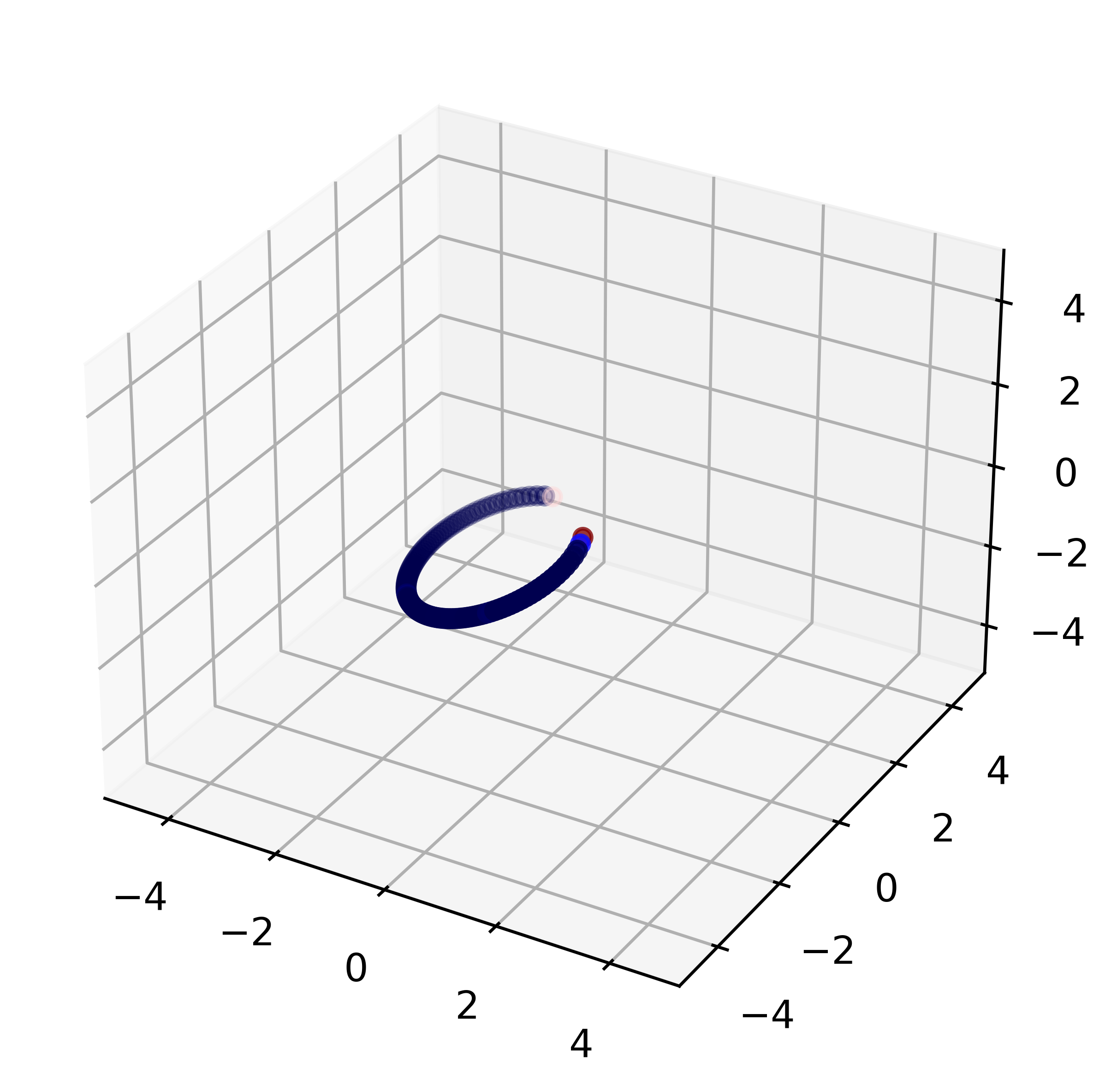}

\includegraphics[width=0.23\linewidth]{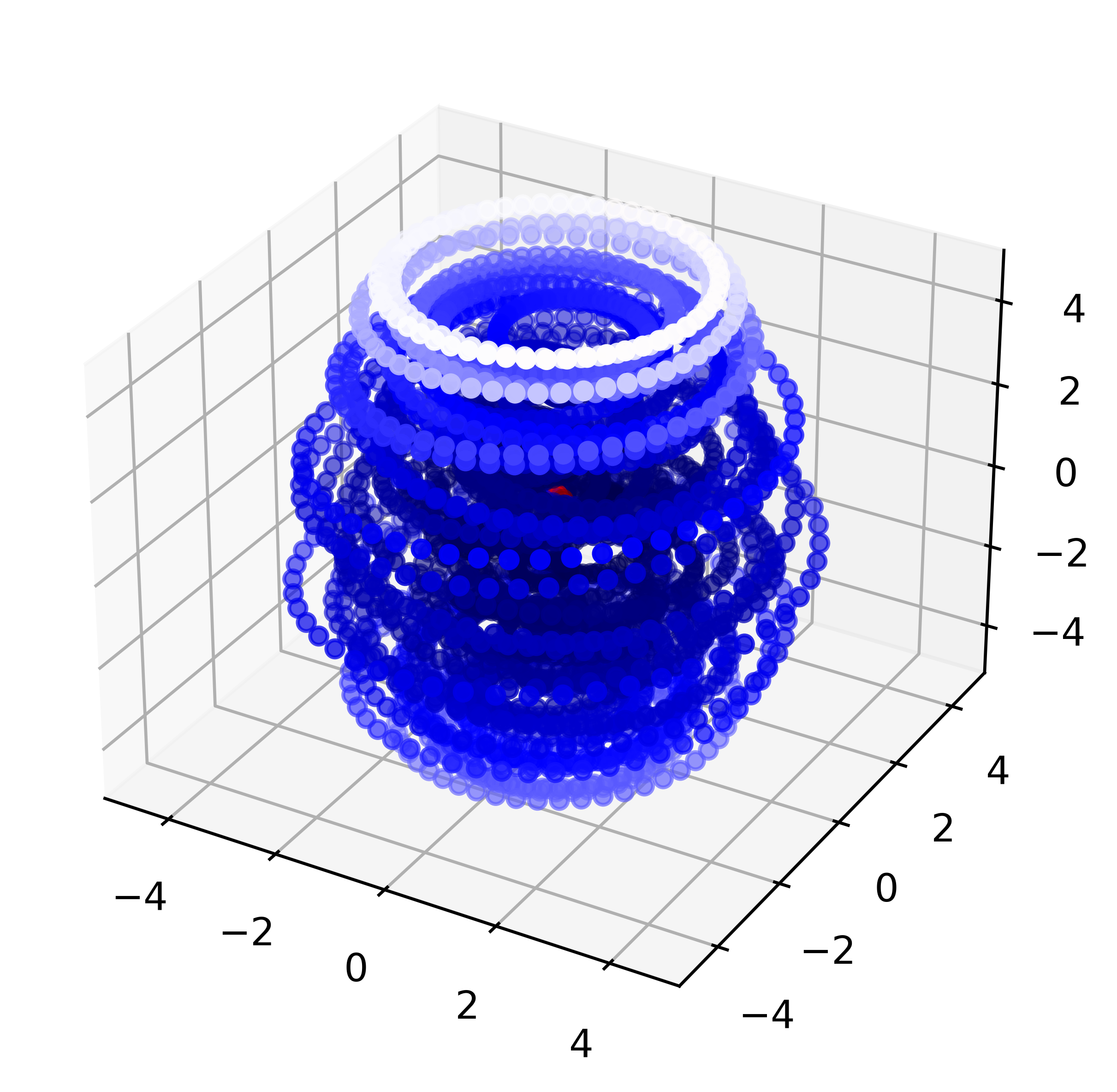}\quad
\includegraphics[width=0.23\linewidth]{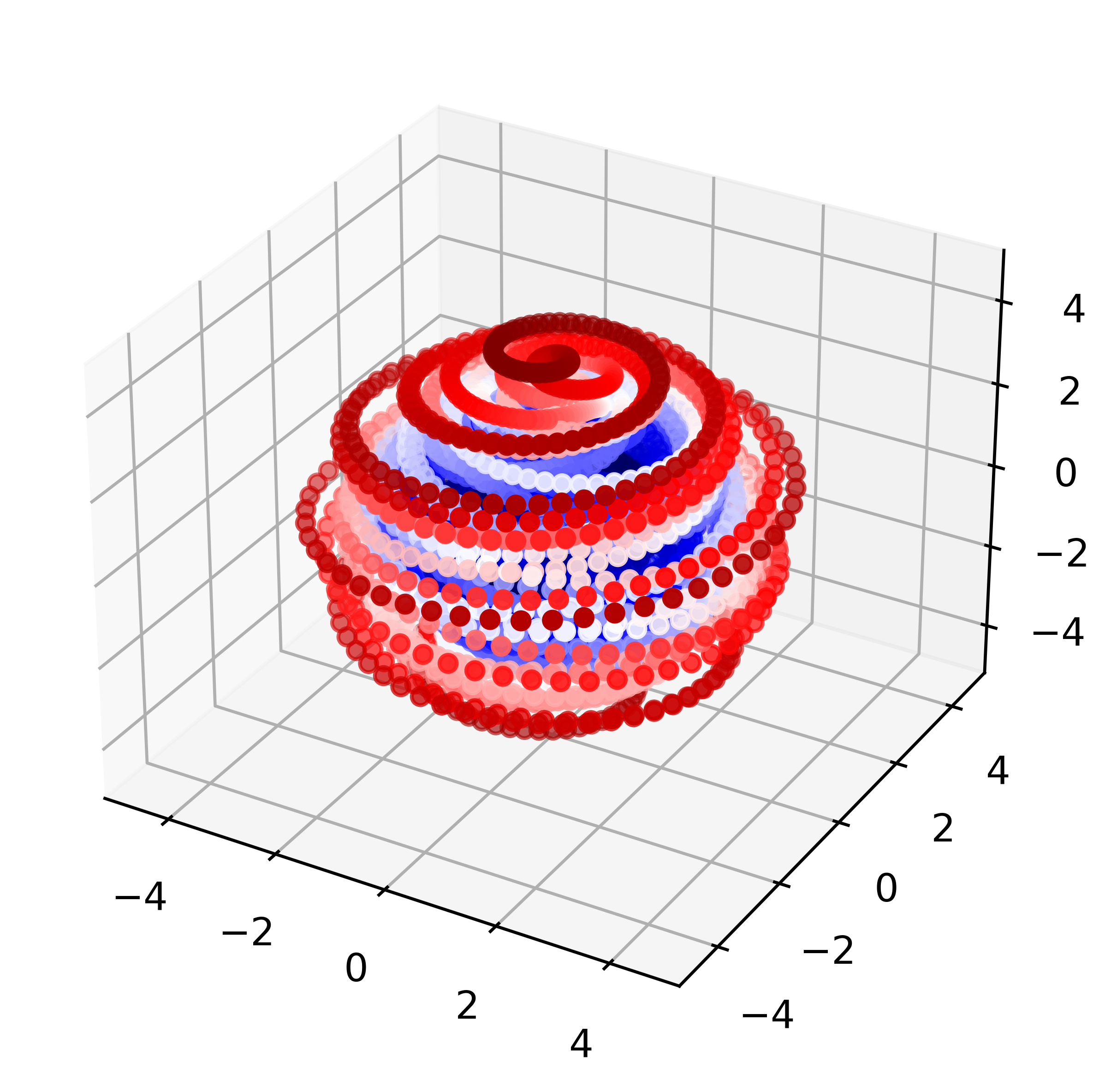}\quad
\includegraphics[width=0.23\linewidth]{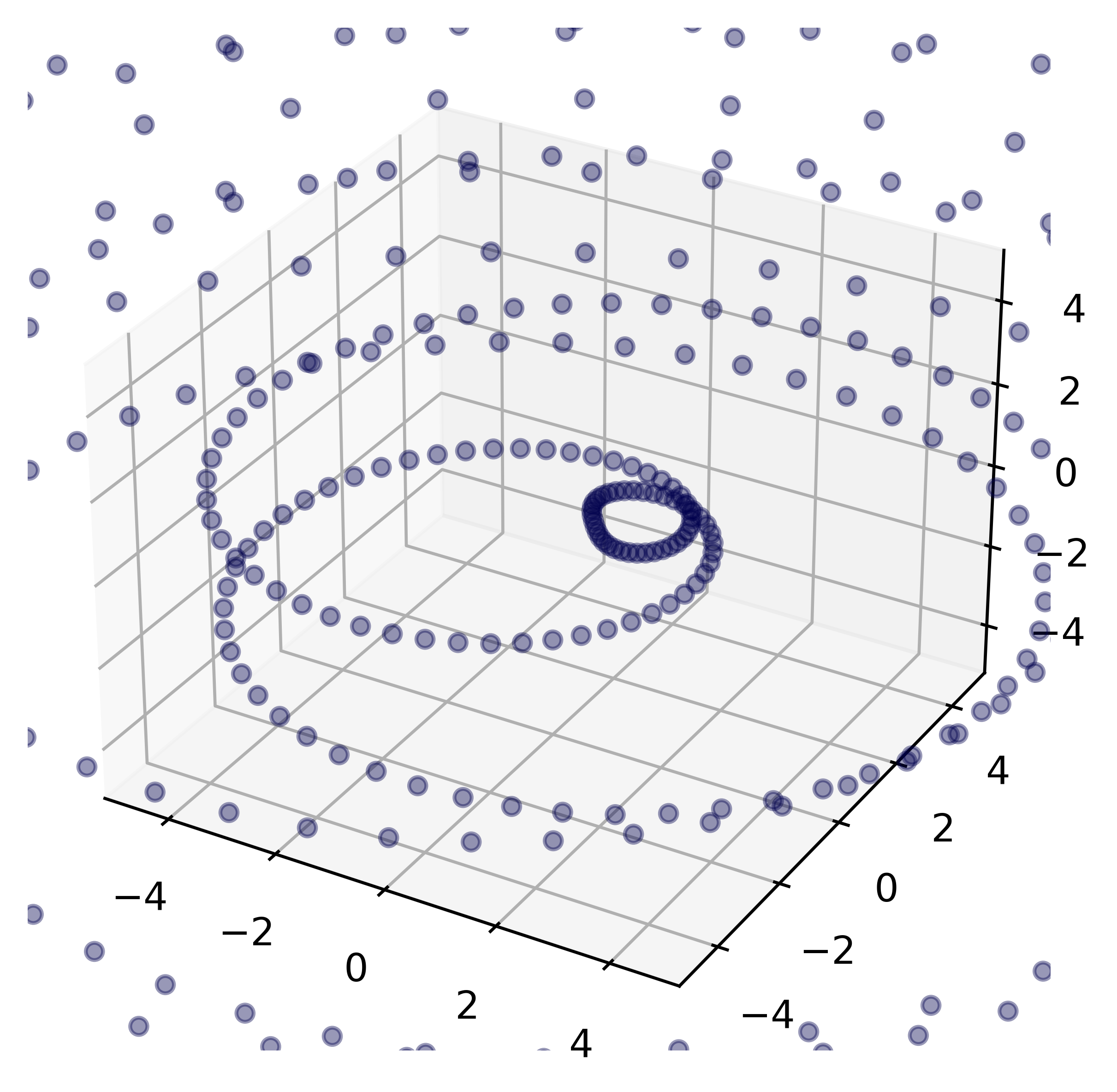}\quad
\includegraphics[width=0.23\linewidth]{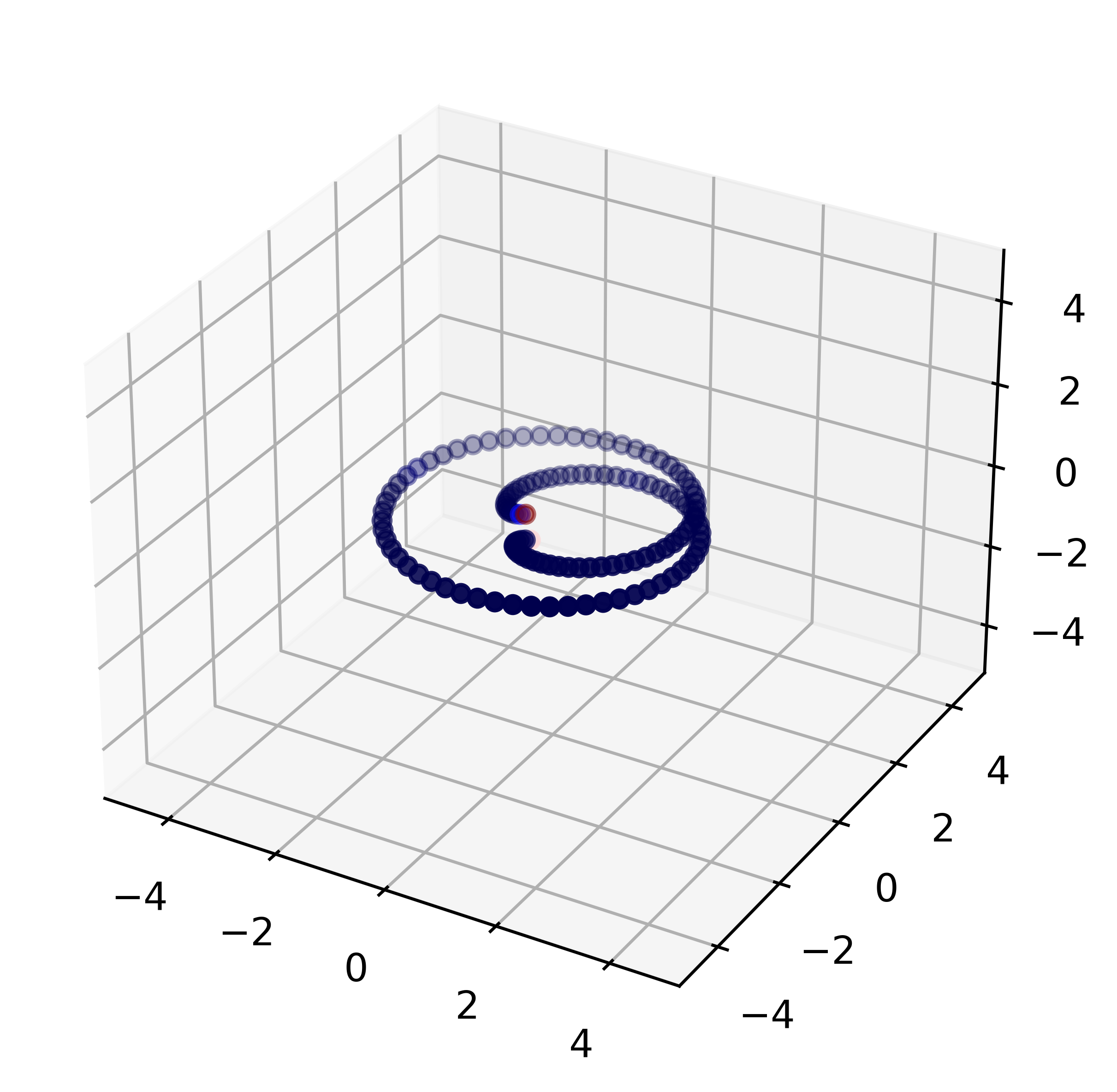}
\caption{Trajectories of objects with Spice IDs -64000001, 64000102, 64000164 and 6400002 in J2000 inertial frame (top) and body-fixed frame around Bennu (bottom). Color signifies the difference between the measured acceleration and the acceleration expected from a homogeneous mass model. Low discrepancy is colored in blue, high discrepancy in red.}
\label{fig:example-pebbles}
\end{figure}

\begin{figure}
\centering
\subcaptionbox{Ratio of magnitudes of measured acceleration vs. accelerations predicted from a homogenous mass distribution (mascon model). \label{fig:relative-mascon-error}}
{\includegraphics[width=0.45\linewidth]{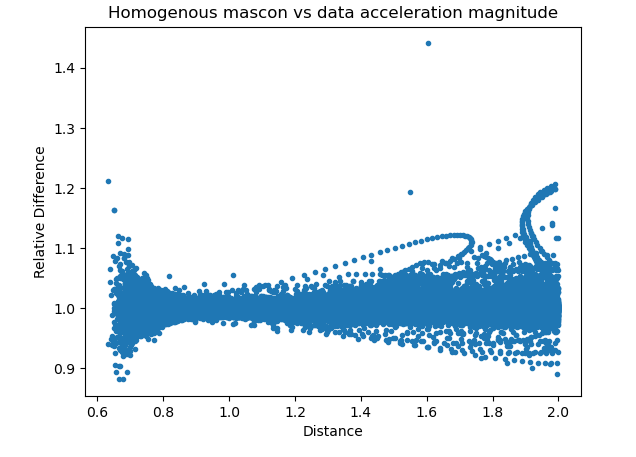}}
\subcaptionbox{Histogram of distances among sampled pebble positions. \label{fig:distance-histogram}}
{\includegraphics[width=0.45\linewidth]{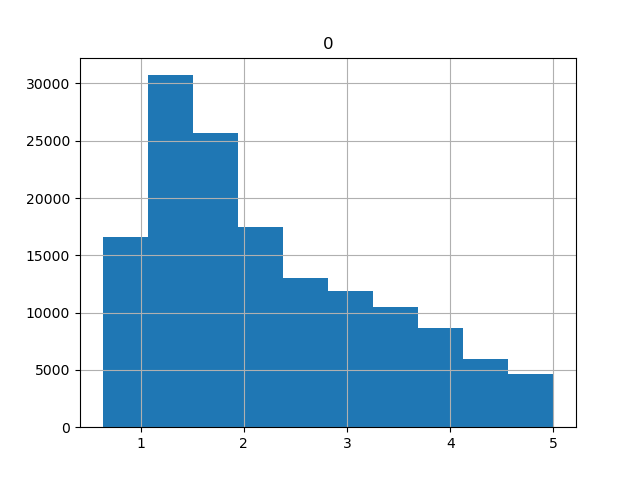}}
\caption{Properties of dataset sampled from pebbles at 5 minute intervals}
\label{fig:mascon-error-and-distance}
\end{figure}

The ratio of the gravitational acceleration and other forces, for example solar radiation pressure, depends strongly on the distance to Bennu's center of mass. Overall, the acceleration measurements of the pebble dataset are considerably more noisy than the one gathered from Osiris Rex, not alone because the precise shape and size of the pebbles, and thus their balance of forces, is unknown. Figure~\ref{fig:example-pebbles} shows the trajectories of four representative pebbles around Bennu. The colors show the relative difference to the acceleration expected from a homogeneous mass model. It can be seen that this difference grows with increasing distance from Bennu, though this is possibly because constant uncertainties in modeling the solar radiation pressure are relatively larger when the gravitational force is weaker. Figure~\ref{fig:relative-mascon-error} shows the ratios of acceleration magnitude between the measured accelerations and the predicted accelerations from a homogeneous shape model.

\subsection{Processing}
For use in neural network training, we scale Bennu to fit into a [-0.8, 0.8] cube and adapt the other units to yield a total mass of 1. This results in length units of 352.1486930549145 m, time units of 2987.9522951342647 s and acceleration units of 3.94e-05 $m/s^2$. All points are rotated into a body-fixed frame around Bennu, for which we use rotation matrices supplied by Spice.

To exclude data points which experience large non-gravity forces, we apply different filters:
\begin{enumerate}
    \item \emph{Distance filters}, as the relative influence of non-gravity forces increases both close and far away from Bennu.
    \item \emph{Angle filters}: When the acceleration vector after subtracting solar radiation pressure points too far away from Bennu's center of mass, it indicates non-gravity forces rather than heterogeneity.
\end{enumerate}

\subsection{Mascon Model}
For the differential scenario and as a comparison, we consider the digital shape model of Bennu provided by the Osiris Rex mission\footnote{\url{https://www.asteroidmission.org/updated-bennu-shape-model-3d-files/}} and as a SPICE Digital Shape Kernel (DSK). This is a triangular surface model, which we convert into a tetrahedral representation with the tool \emph{tetgen}~\cite{tetgen}. In the volumetric center of each tetrahedron, we place a mass point with a mass proportional to its volume. We normalize the complete mass to 1.

\section{EXPERIMENTS}
\label{sec:experiments}

\subsection{Non-differential: Inverting the Topology}

We first consider the scenario where no shape model of Bennu is available and set out to infer the asteroid topology solely from its gravitational field.
For that, we sample the 313 known pebble trajectories in intervals of five minutes, resulting in \numprint{345048} samples.
From these, we select the \numprint{100277} samples which are at most 1000 meters and at least 290 meters away from Bennu's center of mass. (The latter filter is necessary because due to numerical issues, some of the reported trajectories intersect with the asteroid's surface.)
In line with assuming a limited knowledge of Bennu's mass distribution, we omit filters designed for outlier detection that make use of a shape model. We use a train-test split of 80\% training samples, 20\% validation samples and train for \numprint{10000} iterations with a batch size of 100.

\begin{figure}
\centering
\subcaptionbox{Non-differential model \label{fig:inverted-shape-non-differential}}
{\includegraphics[width=0.45\linewidth]{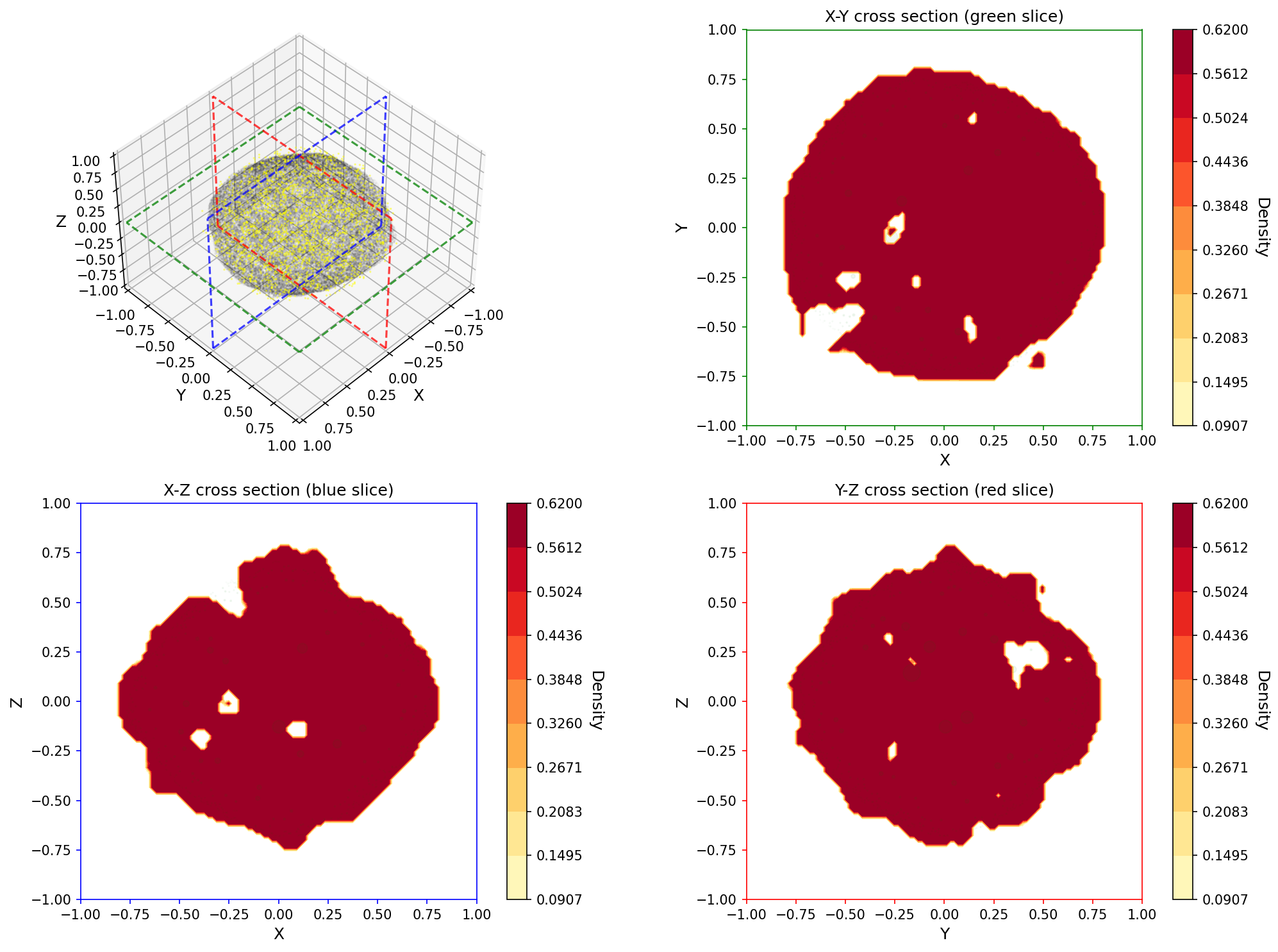}}
\subcaptionbox{Differential model \label{fig:density-differential}}
{\includegraphics[width=0.45\linewidth]{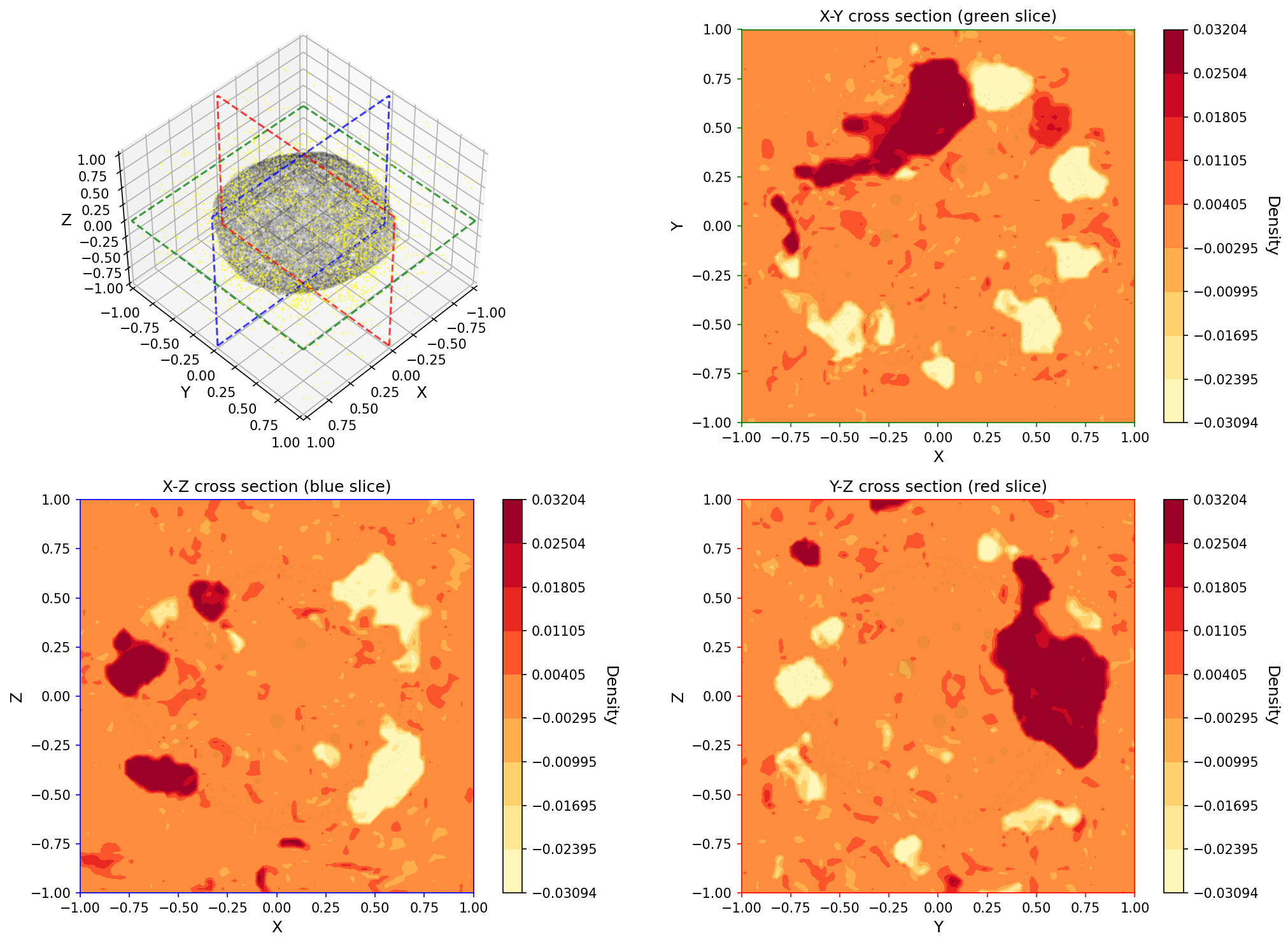}}
\caption{Heat maps of neural density models trained on pebble trajectories. Note that the scale for the differential model is much smaller.}
\end{figure}

Figure~\ref{fig:inverted-shape-non-differential} shows a predicted mass density of such a run.
As the topology inversion problem is ill-posed and multiple topologies result in the same gravitational field.
Thus, each training run results in \emph{a} mass density, targeted to be compatible with the observed field. 
However, in all of them the shape of Bennu is roughly visible.
Note that the lower density of the equatorial bulge was not recovered, instead in this example multiple caves are predicted in the equatorial region, likely the effect of the non gravitational noise present in our data.

\subsubsection{Orbit propagation}

To examine the fidelity of the modeled gravitational field, we choose a set of pebbles with long observable durations, remove them from our training dataset, retrain the network and propagate the trajectories of the holdout pebbles based on the gravity field predicted by the network. For this, we choose pebbles that have at least two days of observable data and do not pass beyond 100km of distance, as after that gravitational forces become negligible and modeling uncertainty in solar pressure radiation dominates. We repeat the training five times to average out randomness in network training and initialization, then choose the model with the lowest validation error. %TODO: we also exclude those that throw Spice Insufficient Data Exceptions when trying to propagate them, but I don't know how to describe this.

\begin{figure}
\centering
\subcaptionbox{Non-differential model \label{fig:prop-error-non-diff}}
{\includegraphics[width=0.45\linewidth]{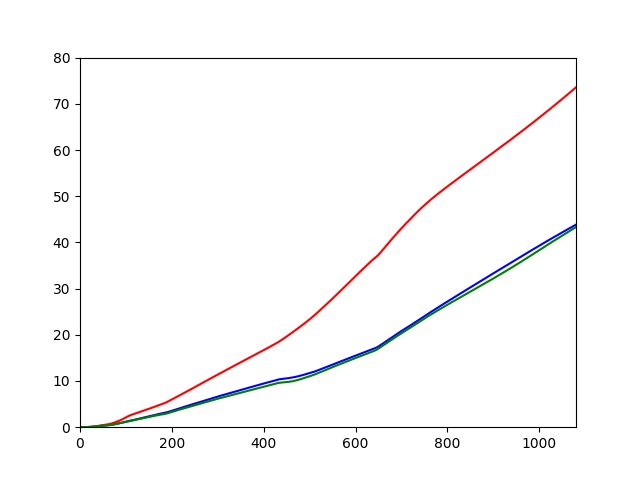}}
\subcaptionbox{Differential model \label{fig:prop-error-diff}}
{\includegraphics[width=0.45\linewidth]{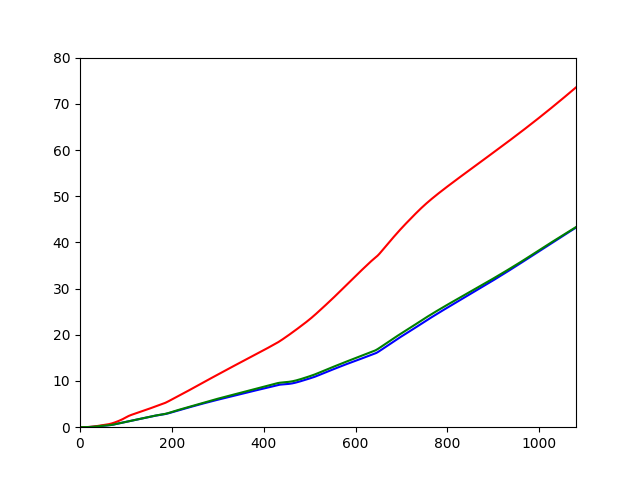}}
\caption{Average propagation errors for 1080 timesteps of 10 seconds each, corresponding to three hours. Units on the x axis are timesteps, units on the y axis are meters. Propagation using the neural network accelerations is shown in blue, mascon model of uniform density is shown in green, point-mass model is shown in red. The error between the network and the mascon model is comparable, while the error caused by the point-mass model is much higher.}
\end{figure}

After completing the training phase, we propagate each pebble in time steps of 10 seconds each, comparing the accelerations from our network, a point-mass model and a mascon model of uniform density. For each, we add solar radiation pressure as calculated in Section\ref{sec:srp}. Figure~\ref{fig:prop-error-non-diff} shows a plot of the resulting deviations from the ground-truth trajectories. After propagating for \numprint{10800} seconds (3 hours), the average distance from the ground-truth trajectories is 44.47 meters for the accelerations learned from the neural network, 73.47 meters when modeling Bennu as a point mass, and 43.26 meters with a mascon model assuming uniform density. While this deviation is considerably large, it still shows that our approach learned the gravitational field to a similar fidelity as a mascon model, even in the presence of large non-gravity forces and without requiring a shape model.

\subsection{Differential: Mass Density Variations}

In the differential case, we assume to have a shape model available and we try to infer the difference to a homogeneous mass density model.

The data point selection and training regime mirror those of the non-differential case, an illustration of the resulting differential density is shown in Figure~\ref{fig:density-differential}.
As this approach aims to model the differences to a uniform mass density, the predicted (and illustrated) density contains both positive and negative components. We find that its magnitude, however, is much smaller and thus the differential neural density field vanishes indicating that we are unable to learn from our data about the actual internal heterogeneity of Bennu. The same conclusion can be drawn looking at the actual trajectory propagation errors, see Figure~\ref{fig:prop-error-diff}.

We assume this is because the heterogeneity of Bennu is small compared to the modeling uncertainty in pebble sizes and resulting non-gravitational forces, which causes the normalization factor $\kappa$ (see \eqref{eq:network-loss}) to vanish.

\subsection{Sample Location and Nuisance Forces}
\begin{figure}
    \centering
    \includegraphics[width=0.45\linewidth]{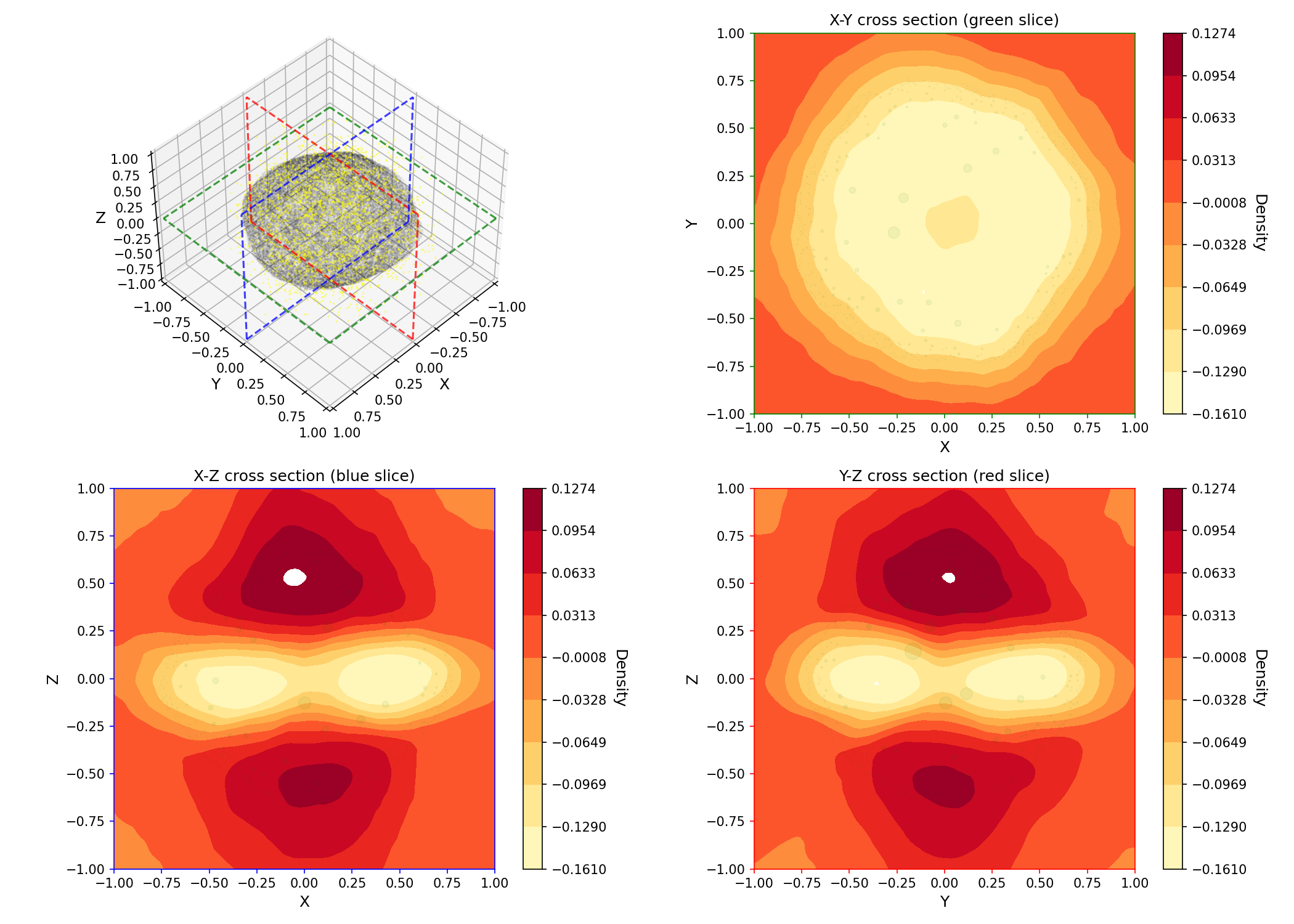}
    \caption{Neural density inferred from observed pebble positions, but with synthetic accelerations of a non-uniform mascon model with a lower density at the equator.}
    \label{fig:pebble-pos-with-mascon-nu}
\end{figure}

The synthetic data used in our companion paper~\cite{izzo2021geodesy} have two aspects making them more convenient for gravity-based geodesy: A uniform coverage of all longitudes and latitudes at close distance, and the absence of non-gravity forces. To investigate which of these differences matters more, we train a network using the positions sampled from the pebble trajectories as in Section~\ref{sec:data}, but with synthetic accelerations generated with a non-uniform mascon model. The resulting density, shown in Figure~\ref{fig:pebble-pos-with-mascon-nu}, captures the artificial heterogeneity clearly, implying that an inversion of the gravity field can be achieved even with non-uniform sample positions, as long as all non-gravity forces are accounted for.

\subsection{Neural Networks}

The concept of geodesyNets takes inspiration from a number of recent works at the interface of computer vision and artificial intelligence with the most notable one being the work on so-called Neural Radiance Fields (NeRF) by Mildenhall \etal{} \cite{mildenhall2020nerf}. In their work they use NeRFs to reconstruct three-dimensional objects from a set of two-dimensional images. Their training setup relies on the network rendering unseen image angles which are then directly compared to real images, similar to the training in geodesyNets. Beyond these works, there is a larger research corpus exploring approaches utilizing neural networks in the solution of inverse problems \cite{padmanabha2021solving,xu2019neural,kim2020inverse,gomez2018laryngeal}.

\section{Conclusion}

We applied geodesyNets to learn the mass density distribution of the asteroid Bennu from the observed trajectories of particles (pebbles) orbiting the asteroid Bennu. In the non-differential case, we achieved a performance comparable to modeling Bennu with a mascon model assuming a uniform density, but without requiring a shape model. This conclusion was validated by comparing the propagation of trajectories in the modeled gravity fields.

Unlike the results obtained using synthetic data (see our companion paper~\cite{izzo2021geodesy}) we were not able to improve the fidelity further by applying a differential approach. It seems likely that this is because the uncertainty in modeling the exact size and mass of the pebbles, and thus the experienced acceleration from solar radiation pressure, introduces a bigger effect with respect to the mass density heterogeneity of Bennu.

Extending our modeling of the observed gravitational accelerations to include pebble sizes is thus a natural part of future work. Another interesting avenue is combining this work with approaches to perform neural network operations on-board of spacecraft, thus contributing to the mission autonomy.
%- On-board inversion

\bibliographystyle{IEEEtran}
\bibliography{references}

\end{document}